\documentclass[ preprint,  amsmath,amssymb,  aps, ]{revtex4-1}%
\usepackage{graphicx}
\usepackage{dcolumn}
\usepackage{bm}
\usepackage{amsmath}
\usepackage{amsfonts}
\usepackage{amssymb}%
\providecommand{\U}[1]{\protect\rule{.1in}{.1in}}

\begin{document}


\title{Detecting quadrupole: a hidden source of magnetic anisotropy for Manganese alloys}%

\author{Jun Okabayashi$^{1}$\footnote{Corresponding author, e-mail address; jun@chem.s.u-tokyo.ac.jp}, Yoshio Miura$^2$, Yohei Kota$^3$, 
Kazuya Suzuki$^{4,5}$, Akimasa Sakuma$^{6,5,7}$, and Shigemi Mizukami$^{4,5,7}$}%

\affiliation{
$^1$Research Center for Spectrochemistry, The University of Tokyo, 113-0033 Tokyo, Japan
}
\affiliation{
$^2$Research Center for Magnetic and Spintronic Materials, National Institute for Materials Science (NIMS), Tsukuba 305-0047, Japan
}
\affiliation{
$^3$National Institute of Technology, Fukushima Collage, Iwaki, Fukushima 970-8034, Japan
}
\affiliation{
$^4$WPI-Advanced Institute for Materials Research, Tohoku University, Sendai 980-8577, Japan
}
\affiliation{
$^5$Center for Spintronics Research Network (CSRN), Tohoku University, Sendai 980-8579, Japan
}
\affiliation{
$^6$Department of Applied Physics, Tohoku University, Sendai 980-8579, Japan
}
\affiliation{
$^7$Center for Science and Innovation in Spintronics (CSIS), Tohoku University, Sendai 980-8577, Japan
}

\date{\today}%
%

\begin{abstract}%

\bf{
Mn-based alloys exhibit unique properties in the spintronics materials possessing perpendicular magnetic anisotropy (PMA) beyond the Fe and Co-based alloys. It is desired to figure out the quantum physics of PMA inherent to Mn-based alloys, which have never been reported. Here, the origin of PMA in ferrimagnetic Mn$_{3-{\delta}}$Ga ordered alloys is investigated to resolve antiparallel-coupled Mn sites using x-ray magnetic circular and linear dichroism (XMCD/XMLD) and a first-principles calculation. We found that the contribution of orbital magnetic moments in PMA is small from XMCD and that the finite quadrupole-like orbital distortion through spin-flipped electron hopping is dominant from XMLD and theoretical calculations. These findings suggest that the spin-flipped orbital quadrupole formations originate from the PMA in Mn$_{3-{\delta}}$Ga and bring the paradigm shift in the researches of  PMA materials using x-ray magnetic spectroscopies. 
}
\end{abstract}%

\maketitle

\section*{\label{sec:level1}Introduction}

Perpendicular magnetic anisotropy (PMA) is desired for the development of high-density magnetic storage technologies. Thermal stability of ultrahigh density magnetic devices is required to overcome the superparamagnetic limit$^{1-3}$. Recently, research interests using PMA films have focused on not only magnetic tunnel junctions$^{4-7}$ toward the realization of spin-transfer switching magneto-resistive random-access memories but also antiferromagnetic or ferrimagnetic devices$^{8,9}$. To design PMA materials, heavy-metal elements that possess large spin-orbit coupling are often utilized through the interplay between the spins in 3$d$ transition-metals (TMs) and 4$d$ or 5$d$ TMs. The design of PMA materials without using the heavy-metal elements is an important subject in future spintronics researches. Recent progress has focused on the interfacial PMA in CoFeB/MgO$^{10}$ or Fe/MgO$^{11,12}$. However, a high PMA of over the order of MJ/m$^3$ with a large coercive field is needed to maintain the magnetic directions during device operation$^{13}$. Therefore, the materials using high PMA constants and without using heavy-metal atoms are strongly desired.

Mn-Ga binary alloys are a candidate that could overcome these issues. Mn$_{3-{\delta}}$Ga alloys satisfy the conditions of high spin polarization, low saturation magnetization, and low magnetic damping constants$^{14-18}$. Tetragonal Mn$_{3-{\delta}}$Ga alloys are widely recognized as hard magnets, which exhibit high PMA, ferromagnetic, or ferrimagnetic properties depending on the Mn composition$^{15}$. Two kinds of Mn sites, which couple antiferromagnetically, consist of Mn$_{3-{\delta}}$Ga with the $D0_{22}$-type ordering. Meanwhile, the $L1_0$-type Mn$_1$Ga ordered alloy possesses a single Mn site. These specific crystalline structures provide the elongated $c$-axis direction, which induces the anisotropic chemical bonding, resulting in the anisotropy of electron occupancies in 3$d$ states and charge distribution. There are many reports investigating the electronic and magnetic structures of Mn$_{3-{\delta}}$Ga alloys to clarify the origin of large PMA and coercive field$^{19-21}$. To investigate the mechanism of PMA and large coercive fields in Mn$_{3-{\delta}}$Ga, site-specific magnetic properties must be investigated explicitly.

X-ray magnetic circular/ linear dichroism (XMCD/ XMLD) may be a powerful tool to study the orbital magnetic moments and magnetic dipole moments of higher order term of spin magnetic moments$^{22,23}$. However, the difficulty in the deconvolution of two kinds of Mn sites prevents site-selected detailed investigations. Within the magneto-optical spin sum rule, using the formulation proposed by C.T. Chen $et$ $al$.$^{24}$, the orbital magnetic moments are expressed as proportional to $r/q$, where $q$ and $r$ represent the integral of the x-ray absorption spectra (XAS) and XMCD spectra, respectively, for both $L_2$ and $L_3$ edges. In the cases of two existing components, the orbital moments are not obtained from the whole integrals of spectra; by using each component $r_1$, $r_2$, $q_1$, and $q_2$, the value of $(r_1/q_1)$ + $(r_2/q_2)$ should be the average value. The value of ($r_1+r_2$)/($q_1+q_2$) does not make sense as an average in the case of core-level atomic excitation, leading to the wrong value in the XMCD analysis. As a typical example, for the mixed valence compound CoFe$_2$O$_4$, the Fe$^{3+}$ and Fe$^{2+}$ sites can be deconvoluted by the ligand-field theory approximation$^{25}$. However, the deconvolution of featureless line shapes in a metallic Mn$_{3-{\delta}}$Ga case is difficult by comparison with the theoretical calculations. To detect the site-specific anti-parallel-coupled two Mn sites, systematic investigations using Mn$_{3-{\delta}}$Ga of ${\delta}$ = 0, 1, and 2 provide the information of site-specific detections. In previous reports, although Rode $et$ $al.$ performed XMCD measurements of Mn$_3$Ga and Mn$_2$Ga$^{19}$, the comparison of XMCD line shapes with that of the single Mn sites in Mn$_1$Ga is necessary. Recently, the crystal growth techniques of Mn$_{3-{\delta}}$Ga were developed using CoGa buffer layers to synthesize composition-controlled Mn$_{3-{\delta}}$Ga thin films$^{26}$, which enables discussion of the electronic structures of Mn$_{3-{\delta}}$Ga in ${\delta}$ = 0, 1, and 2. We adopted this growth technique and performed XMCD. By contrast, the XMLD in Mn $L$-edges enables detection of the element of quadrupole tensor $Q_{zz}$ by adopting the XMLD sum rule$^{27}$. Although many reports of XMLD for in-plane magnetic easy axis cases exist, perpendicularly magnetized cases are attempted firstly in Mn$_{3-{\delta}}$Ga using perpendicular remnant magnetization states.

The first-principles calculations based on the density functional theory (DFT) suggest the unique band structures in the Mn sites of mixed spin-up and -down bands at the Fermi level ($E_\mathrm{F}$), which allow the spin transition between up and down spin states, resulting in the stabilization of the PMA$^{28-30}$. Usually, the PMA originates from the anisotropy of orbital magnetic moments in the large exchange-split cases, such as Fe and Co, using second-order perturbation for spin-orbit interaction$^{31,32}$. Meanwhile, for Mn compounds, the contribution from orbital moment anisotropy for PMA is smaller than the spin-flipped contribution to PMA$^{33-35}$. However, this picture has not been guaranteed completely from an experimental viewpoint until now.

In this study, we performed the deconvolution of each Mn site using the systematic XMCD and XMLD measurements for different Mn contents in Mn$_{3-{\delta}}$Ga. We discuss the site-specific spin and orbital magnetic moments with magnetic dipoleterm, which corresponds to electric quadrupoles. These are deduced from the angular-dependent XMCD and XMLD and compared with the DFT calculations to understand the PMA microscopically.

\section*{\label{sec:level1}Results}
\subsection*{\label{sec:level1}X-ray magnetic spectroscopies}
The Mn $L$-edge XAS and XMCD for L1$_0$-type Mn$_1$Ga with a single Mn site (MnI), and D0$_{22}$-type Mn$_2$Ga and Mn$_3$Ga with two kinds of Mn sites (MnI and MnII) are shown in Fig. 1. The XAS were normalized to be one at the post-edges. With increasing Mn concentrations (decreasing ${\delta}$), the intensities of XAS increased and the difference between ${\mu}^+$ and ${\mu}^-$ became small, resulting in the suppression of XMCD intensities because of the increase of antiparallel components. In the case of Mn$_2$Ga and Mn$_3$Ga, the XMCD line shapes in the $L_3$ and $L_2$ edges, of slightly split and doublet structures, became clear because of the increase of another MnII component with opposite sign. Furthermore, the element-specific hysteresis curves in XMCD at a fixed photon energy of Mn $L_3$-edge exhibit similar features with the results of the magneto-optical Kerr effects. Coercive fields ($H_\mathrm{c}$) of 0.5 T were obtained for the Mn$_2$Ga and Mn$_3$Ga cases because the two kinds of Mn sites enhance the antiparallel coupling.

To deconvolute the MnI and MnII sites in the XMCD spectra, we performed the subtraction of XMCD between Mn$_1$Ga and Mn$_3$Ga. Figure 2 displays the XMCD of Mn$_1$Ga and Mn$_3$Ga, and their differences after the normalization considering the Mn compositions. The XMCD signal with opposite sign was clearly detected for MnI and MnII components. As the lattice volume of Mn$_{3-{\delta}}$Ga on the CoGa buffer layer remained almost unchanged with different $\delta$ $^{26}$, the validity of the subtraction of XMCD is warranted because the density of states (DOS) for MnI is similar in all ${\delta}$ regions as shown in latter (Fig. 4). To apply the magneto-optical sum rule for effective spin magnetic moments ($m_s^{\mathrm{eff}}$) including magnetic dipole term and orbital magnetic moments ($m_\mathrm{orb}$), the integrals of the XMCD line shapes are needed$^{24}$. Further, the integrals of XAS were also estimated for MnI and MnII, divided by the composition ratios. The electron numbers for 3$d$ states of MnI and MnII were estimated from the band-structure calculations to be 5.795 and 5.833, respectively. Thus, $m_s^\mathrm{eff}$ and $m_\mathrm{orb}$ for MnI were estimated to be 2.30 and 0.163 ${\mu}_\mathrm{B}$, respectively. For MnII, 2.94 ${\mu}_\mathrm{B}$ ($m_s^\mathrm{eff}$) and 0.093 ${\mu}_\mathrm{B}$ ($m_\mathrm{orb}^{\perp}$) were obtained for perpendicular components with the error bars of 20 \% because of the ambiguities estimating spectral background. 

Here, we claim the validity of $m_s^\mathrm{eff}$ and $m_\mathrm{orb}$ in Mn$_{3-{\delta}}$Ga deduced from XMCD. First, these $m_\mathrm{orb}$ values are too small to explain stabilizing the PMA because the magnetic crystalline energy  $E_\mathrm{MCA}{\propto}{\frac{1}{4}}{\alpha}{\xi}_\mathrm{Mn}$ $(m_{\mathrm{orb}}^{\perp}-m_{\mathrm{orb}}^{\parallel}) $ 
within the scheme of the Bruno relation$^{32}$, assuming the spin-orbit coupling constant  $\xi_\mathrm{Mn}$ of 41 meV and the band-state parameter ${\alpha}=0.2$ for Mn compounds, which is estimated from the DFT calculation. For Mn$_1$Ga, as the saturation magnetic field along hard axis direction was less than 1 T, the projected component $m_\mathrm{orb}^{\parallel}$ could be deduced as ${\Delta}m_\mathrm{orb}(=m_{\mathrm{orb}}^{\perp}-m_{\mathrm{orb}}^{\parallel}) $ of less than 0.01 ${\mu}_\mathrm{B}$, resulting in  $E_\mathrm{MCA}=1{\times}10^{-5}$ eV/atom, that is, 5.7 $\times$ 10$^4$ J/m$^3$ using the unit cell of MnGa. Therefore, orbital moment anisotropy cannot explain the PMA of the order of 10$^6$ J/m$^3$ in Mn$_{3-{\delta}}$Ga$^{26}$. As the electron configuration is close to the half-filled 3$d^5$ case, the quenching of the orbital angular momentum occurs in principle. In Mn$_{3-{\delta}}$Ga, since the electron filling is not complete half-filled cases, small orbital angular momentum appears. Second, another origin for the large PMA is considered as the spin-flipped contribution between the spin-up and -down states in the vicinity of the $E_\mathrm{F}$. The magnetic dipole term ($m_\mathrm{T}$) also stabilizes the magneto-crystalline anisotropy energy ($E_\mathrm{MCA}$) by the following equation$^{33}$:
\begin{equation}
E_\mathrm{MCA} {\simeq} {\frac{1}{4}}{\alpha}{\xi}{\Delta}m_\mathrm{orb}-\frac{21}{2}\frac{{\xi}^2}{{\Delta}E_\mathrm{ex}}{\Delta}m_\mathrm{T},
\end{equation}
where ${\Delta}E_\mathrm{ex}$ denotes the exchange splitting of 3$d$ bands. Positive values of $E_\mathrm{MCA}$ stabilize the PMA. The second term becomes dominant when proximity-driven exchange split cases, such as the 4$d$ and 5$d$ states, are dominant$^{36,37}$. In the case of Mn$_{3-{\delta}}$Ga, the Mn 3$d$ states were delicate regarding the mixing of the spin-up and -down states at the  $E_\mathrm{F}$, which corresponds to the quadrupole formation and the band structure ${\alpha}$ values. The second term is expressed by $m_\mathrm{T}$ in the XMCD spin sum rule of $m_s+7m_\mathrm{T_z}$ along the out-of-plane $z$ direction$^{38}$. For Mn$_1$Ga, if $m_{T_z}$ is negative, resulting in $Q_{zz}\verb|>|$ 0 in the notation of $m_\mathrm{Tz}=-Q_{zz}{\cdot}m_s$, which exhibits the prolate shape of the spin density distribution; the second term favors PMA because of the different sign for the contribution of orbital moment anisotropy in the first term. Since 7$m_\mathrm{Tz}$ is estimated to be in the order of 0.1 ${\mu}_{\mathrm{B}}$ from angular-dependent XMCD between surface normal and magic angle cases, $Q_{zz}$ is less than 0.01, resulting that the orbital polarization of less than 1\% contributes to stabilize PMA.   In this case, the contribution of the second term in eq.(1) is one order larger than the orbital term, which is essential for explaining the PMA of Mn$_{3-{\delta}}$Ga.  
Third, in a previous study$^{19}$, quite small ${\Delta}m_\mathrm{orb}$ and negligible $m_\mathrm{T_z}$ were reported for Mn$_2$Ga and Mn$_3$Ga. Their detailed investigation claims that ${\Delta}m_\mathrm{orb}$ of 0.02 ${\mu}_\mathrm{B}$ in MnI site contributes to PMA and MnII site has the opposite sign. These are qualitatively consistent with our results. The difference might be derived from the sample growth conditions and experimental setup. Fourth, the reason why $H_\mathrm{c}$ in Mn$_1$Ga is small can be explained by the $L1_0$-type structure, due to the stacking of the Mn and Ga layers alternately, which weakened the exchange coupling between the Mn layers. Finally, we comment on the XMCD of the Ga $L$-edges. This also exhibits the same sign as the MnI component, suggesting that the induced moments in the Ga sites were derived from the MnI component (Fig. S1), which was substituted by the MnII for Mn$_2$Ga and Mn$_3$Ga.

To determine the effect of $m_\mathrm{T_z}$, we performed XMLD measurements. Figure 3 shows the $\bf{E}$ vector polarization dependent XAS, where $\bf{E}$ is perpendicular and horizontal to the magnetization direction. After magnetizing perpendicular to easy-axis direction by the pulse of 1 T, the XMLD was measured at the remnant states. The XMLD between the vertical and horizontal polarized excitations were detected in grazing incident beams, where the sample surface normal is tilted 60$^{\circ}$ from the incident beam. The differential line shapes were similar to those of other Mn compounds$^{39,40}$. We estimate that the x-ray linear dichroism (XLD) components are less than 20 \% by the measurements at the same geometry without magnetizing. With increasing Mn composition, the XMLD signal intensities were enhanced because XMLD detects the ${\langle}M^2\rangle$ contribution. In Mn$_3$Ga, XMLD includes the summation of both ${\langle}M_\mathrm{MnI}^2\rangle+{\langle}M_\mathrm{MnII}^2\rangle$ contributions. Therefore, the contributions from XLD are not dominant factor in spectral analyses. We note that the integrals of the XMLD line shapes are proportional to $Q_{zz}$ along the sample surface normal direction. We confirmed that the integral converges to a positive value, deducing that the sign of $Q_{zz}$ is positive with the order of 0.01 for both MnI and MnII components by applying the XMLD sum rule in the notation of $m_\mathrm{Tz}=-Q_{zz}{\langle}S{\rangle}$; that is, 3$z^2-r^2$ orbitals are strongly coupled with $\bf{E}$ and are elongated to an easy-axis direction after subtracting the XLD contribution. The detail of estimation of $Q_{zz}$ is explained in the Supplemental Note and ref 41. This value is consistent with that estimated from XMCD spin sum rule. These suggest the orbital polarization of Mn 3$d$ states along $z$-axis direction forming the cigar-type prolate unoccupied orbital orientation. 
Therefore, combining both XMCD and XMLD, the order of $m_\mathrm{Tz}$ can be estimated as two order smaller than the spin moments.

\subsection*{\label{sec:level1}DFT calculation}
Figure 4 shows the DOS of Mn$_1$Ga and Mn$_3$Ga with site and orbital-resolved contributions by the DFT calculation. The contributions for the $E_\mathrm{MCA}$ of each atom and spin transition processes through the second-order perturbation of the spin-orbit interaction are also shown for the MnI and MnII sites using a tetragonal unit cell. 
In the DOS of the MnI and MnII sites, all orbital states were split through exchange interaction. However, exchange splitting was incomplete where complete spin splitting was required, in the Bruno formula$^{32}$, which enabled the transitions by spin mixing between occupied spin-up and unoccupied spin-down states. Four types of spin transition processes occurred between the occupied and unoccupied states, as shown in the bottom panel of Fig. 4. The positive values in the presented bar graphs stabilize the PMA. The 'up-down' process implies a virtual excitation from an occupied up-spin state to an unoccupied down-spin state in the second-order perturbation, which forms the magnetic dipoles. 
While spin conserved transition terms are slightly positive, spin-flipped transition terms show dominant contribution to PMA.
The further orbital-resolved anatomy of the spin-flipped transition revealed the transition between the $yz$ and 3$z^2$ states, which induces the prolate-type spin distribution. The matrix elements of $L_x$ between $m={\pm}1$ and 0 through the transitions of different magnetic quantum numbers $m$ were predominant for the MnI site in Mn$_1$Ga (see Supplemental Materials). As shown in Fig. S2, the matrix elements of $L_x$ between $yz$ and 3$z^2$ in spin-flipped transition (blue bar graph) have large positive values, indicating contribution to the PMA. Meanwhile, for $D0_{22}$-type Mn$_3$Ga, the contributions to MCA energy were different. 
Although the spin-conserved transition terms are enhanced as compared with Mn$_1$Ga,
the spin-flipped contributions were still dominant, which explains the suppression of the orbital moment anisotropy in Mn$_1$Ga. The MnI 3$z^2$ orbitals were located near the $E_\mathrm{F}$ in Mn$_3$Ga in the spin-down states, strongly affecting the appearance of the finite matrix elements. Figure S3 shows the large matrix elements of $L_x$ between $yz$ and 3$z^2$ in the spin-flipped transition of Mn I, which are similar to those of Mn$_1$Ga. The difference in the MnII sites between Mn$_1$Ga and Mn$_3$Ga can be derived from the location of neighboring Mn atoms, which promotes exchange interaction between the MnI and MnII sites.

For Mn$_1$Ga, the orbital moments along the $c$- and $a$-axis, $m_\mathrm{orb}^z$ and $m_\mathrm{orb}^x$, were estimated to be 0.022 and 0.0207 ${\mu}_\mathrm{B}$, respectively, by the DFT calculation. The magnetic dipole moments of $m_\mathrm{T_z}$ and $m_\mathrm{T_x}$ were $-0.039$ and $-0.022$ ${\mu}_\mathrm{B}$, respectively. Using Eq. (1), ${\Delta}m_\mathrm{orb}$ and ${\Delta}m_\mathrm{T}$ were estimated to be $0.014$ ${\mu}_\mathrm{B}$ and $-0.017$ ${\mu}_\mathrm{B}$, respectively. We note that $Q_{zz}$ was estimated to be 0.015 from the DFT calculation, which is similar to the results of the XMLD through the relation of $m_\mathrm{T}=-Q_{ij}{\cdot}S$. These values provide the anisotropic energies of the first and second terms in Eq. (1) as $0.014$ meV and 0.13 meV, respectively, by using ${\Delta}E_\mathrm{ex}$ of 2.3 eV and ${\xi}_\mathrm{Mn}$ of 41 meV for Mn atoms. The amplitude of the spin-flipped term is larger than the orbital moment anisotropy to stabilize the PMA energetically. These estimations are consistent with the values deduced from the XMCD and XMLD analyses. Eq. (1) was modified using the energy difference of each component $i$: ${\Delta}E_i={E_i}^{\perp} -{E_i}^{\parallel}$ and the relation of: $E_\mathrm{MCA}={\Delta}E_{L{\uparrow}}+ {\Delta}E_{L{\downarrow}}+{\Delta}E_\mathrm{T}+{\Delta}E_\mathrm{LS}$, where $E_\mathrm{MCA}$ is expressed by the summation of the orbital parts of spin up and down states (${\Delta}E_{L{\uparrow}}+{\Delta}E_{L{\downarrow}}$), the $m_\mathrm{T}$ term, and the residual of ${\Delta}E_\mathrm{LS}$$^{33}$. We confirmed that electron number dependence mainly obeys the ${\Delta}E_\mathrm{T}$ term (Fig. S4). Therefore, the finite value of $T_z$, which contributes to the second term in Eq. (1), is indispensable for the PMA in MnGa.

\section*{\label{sec:level1}Discussion}

Considering the results of the XMCD, XMLD, and DFT calculation, we discuss the origin of PMA in Mn$_{3-{\delta}}$Ga. As the orbital magnetic moments and their anisotropies are small, the contribution of the first term in Eq. (1) is also small, which is a unique property of Mn alloy compounds and contradicts the cases of Fe and Co compounds exhibiting PMA. Beyond Bruno's formula$^{31}$, the mixing of majority and minority bands in Mn 3$d$ states enables the spin-flipped transition and $Q_{zz}$. However, comparing with the CoPd or FePt cases, where the exchange splitting was induced in the 4$d$ or 5$d$ states, a small ${\xi}_\mathrm{Mn}$ and large ${\Delta}E_\mathrm{ex}$ in the Mn 3$d$ states suppress the contribution of the second term. Large $Q_{zz}$ values were brought by the crystalline distortion accompanied by the anisotropic spin distribution, resulting in the PMA energy of Mn$_{3-{\delta}}$Ga exhibiting a similar order with those in heavy-metal induced magnetic materials. Therefore, the large PMA in Mn$_{3-{\delta}}$Ga originates from the specific band structure of the Mn 3$d$ states, 
where the orbital selection rule for the electron hopping through spin-flipped ${\langle}yz{\uparrow}|L_x|z^2{\downarrow}{\rangle}$  provides the cigar-type spin distribution$^{28}$.
As the spin-flipped term of ${\Delta}E_\mathrm{T}+{\Delta}E_\mathrm{LS}(={\Delta}E_\mathrm{sf})$ for PMA energy, except the orbital contributions, can be written as:
\begin{equation}
{\Delta}E_\mathrm{sf}=\sum_{u{\uparrow}, o{\downarrow}}\frac{{\xi}^2}{{\Delta}E_\mathrm{ex}}[{\langle}u{\uparrow}|{L_x}^2|o{\downarrow}{\rangle}-{\langle}u{\uparrow}|{L_z}^2|o{\downarrow}{\rangle}]    
\end{equation}
the difference between the $L_x^2$ and $L_z^2$ terms through the spin-flipped transitions between the occupied ($o$) to unoccupied ($u$) states is significant for the gain of the PMA energy. The matrix elements of ${\langle}u{\uparrow}|{L_x}^2| o{\downarrow}{\rangle}$ were enhanced in the spin-flipped transition between $yz$ and $z^2$, and those of  ${\langle}u|{L_z}^2| o{\rangle}$ were enhanced in the spin-conserved case between $xy$ and $x^2-y^2$ $^{28}$. These transitions favor the magnetic dipole moments of prolate shapes (${\langle}Q_{zz}{\rangle}={\langle}3L_z^2-L^2{\rangle}\verb|>|0$) described by the Mn 3$d$ each orbital angular momenta. We emphasize that the signs of ${\Delta}m_\mathrm{orb}$ and $Q_{zz}$ are opposite, which is essential to stabilize the PMA by the contribution of the second term in Eq. (1). The PMA energy of FePt exhibits around MJ/m$^3$ and the contribution of the second term in Pt is four times larger than the Fe orbital anisotropy energy$^{28}$. Therefore, MnGa has a specific band structure by crystalline anisotropy elongated to the $c$-axis and intra-Coulomb interaction in Mn sites to enhance the PMA without using heavy-metal atoms.


In conclusion, we investigated the origin of PMA in Mn$_{3-{\delta}}$Ga by decomposing into two kinds of Mn sites for XMCD, XMLD, and the DFT calculation. The contribution of the orbital moment anisotropy in Mn$_3$Ga is small and that of the mixing between the Mn 3$d$ up and down states is significant for PMA, resulting in the spin-flipped process through the electron hopping between finite unforbidden orbital symmetries in the 3$d$ states through the quadratic contribution. Composition dependence reveals that the orbital magnetic moments of the two antiparallel-coupled components in Mn sites were too small to explain the PMA. These results suggest that the quadrupole-like spin-flipped states through the anisotropic $L1_0$ and $D0_{22}$ crystalline symmetries are originated to the PMA in Mn$_{3-{\delta}}$Ga. The present study provides a promising strategy to investigate quadrupoles in antiferro or ferrimagnetic materials with PMA.

\section*{\label{sec:level1}MATERIALS AND METHODS}

\subsection*{\label{sec:level1}Sample growth and characterization}
The samples were prepared by magnetron sputtering. The 40-nm-thick Cr buffer layers were deposited on single-crystal MgO (001) substrates at room temperature (RT), and in situ annealing at 700 $^{\circ}$C was performed. Subsequently, a 30-nm-thick Co$_{55}$Ga$_{45}$ buffer layer was grown at RT with $in$-$situ$ annealing at 500 $^{\circ}$C, then 3-nm-thick Mn$_{3-{\delta}}$Ga layers were grown at RT. Here, the composition of Mn$_{3-{\delta}}$Ga films were controlled by Ar gas pressure during deposition and co-sputtering technique with MnGa and Mn target. Finally, a 2-nm-thick MgO capping layer was deposited. Using the CoGa buffer layer, ultra-thin Mn$_{3-{\delta}}$Ga layer deposition was achieved$^{26}$. Curie temperatures of samples are higher than RT. Magnetic anisotropy energy of Mn$_1$Ga and Mn$_2$Ga was estimated to be approximately 0.13 and 0.9 MJ/m$^3$, respectively, at RT, by using vibrational sample magnetometer and tunnel magnetoresistance curves shown in the Ref. 26.

\subsection*{\label{sec:level1}XMCD and XMLD measurements}
The XMCD and XMLD were performed at BL-7A and 16A in the Photon Factory at the High-Energy Accelerator Research Organization (KEK). For the XMCD measurements, the photon helicity was fixed, and a magnetic field of $\pm$1.2 T was applied parallel to the incident polarized soft X-ray beam, defined as ${\mu}^+$ and ${\mu}^-$ spectra. The total electron yield mode was adopted, and all measurements were performed at room temperature. The XAS and XMCD measurement geometries were set to normal incidence, so that both the photon helicity and the magnetic field were parallel and normal to the surface, enabling measurement of the absorption processes involving the normal components of the spin and orbital angular momenta$^{36}$. In the XMLD measurements, the remnant states magnetized to PMA were adopted. For grazing incident measurements in XMLD and XLD, the angle between incident beam and sample surface normal was kept at 60$^{\circ}$ tilting as shown in the inset of Fig. 3. The direction of the electric field of the incident synchrotron beam $\bf{E}$ was tuned horizontally and vertically with respect to the magnetization $\bf{M}$. We define the sign of XMLD by the subtraction of the ($\bf{M}{\parallel}{\bf{E}}$)$-$(${\bf{M}}{\perp}{\bf{E}}$) spectra$^{35}$.

\subsection*{\label{sec:level1}First-principles calculation}
The first-principles calculations of MCA energies  for Mn$_1$Ga and Mn$_3$Ga were performed using the Vienna $ab$ $initio$ simulation package (VASP).  We calculated the second-order perturbation of the spin-orbit interaction to MCA energies for each atomic site using wave functions in the VASP calculations. Details of the perturbation calculation are described in Ref. 42.  In this paper, we estimated the spin-orbit coupling constant of Mn and Ga atom as 41 and 35 meV, respectively, by the calculation.

\begin{acknowledgments}
This work was partially supported by JSPS KAKENHI (Grant No. 16H06332), and Spintronics Research Network of Japan. Part of research was performed under the support from Toyota Physical and Chemical Research Institute. Parts of the synchrotron radiation experiments were performed under the approval of the Photon Factory Program Advisory Committee, KEK (Nos. 2017G060 and 2019G028). 
\end{acknowledgments}

\section*{\label{sec:level1}Author Contributions}
J. O. and S. M. planned the study. Kz. S. prepared the samples and characterized the properties. J.O. set up the XMCD and XMLD measurement apparatus at Photon Factory and collected and analyzed the data. Y.M. performed the first-principles calculation. Y. K. and A. S. also performed the density-functional-theory calculation. J.O and Y.M. constructed the scenario of quadrupole physics. All authors discussed the results and wrote the manuscript.

\section*{\label{sec:level1}Competing interests}
The authors declare no competing financial interests.

\section*{\label{sec:level1}Data availability}
The data that support the findings of this study are available from the corresponding author upon reasonable request.

\begin{figure}
[ptb]
\begin{center}
\includegraphics[
height=8in,
width=6in
]%
{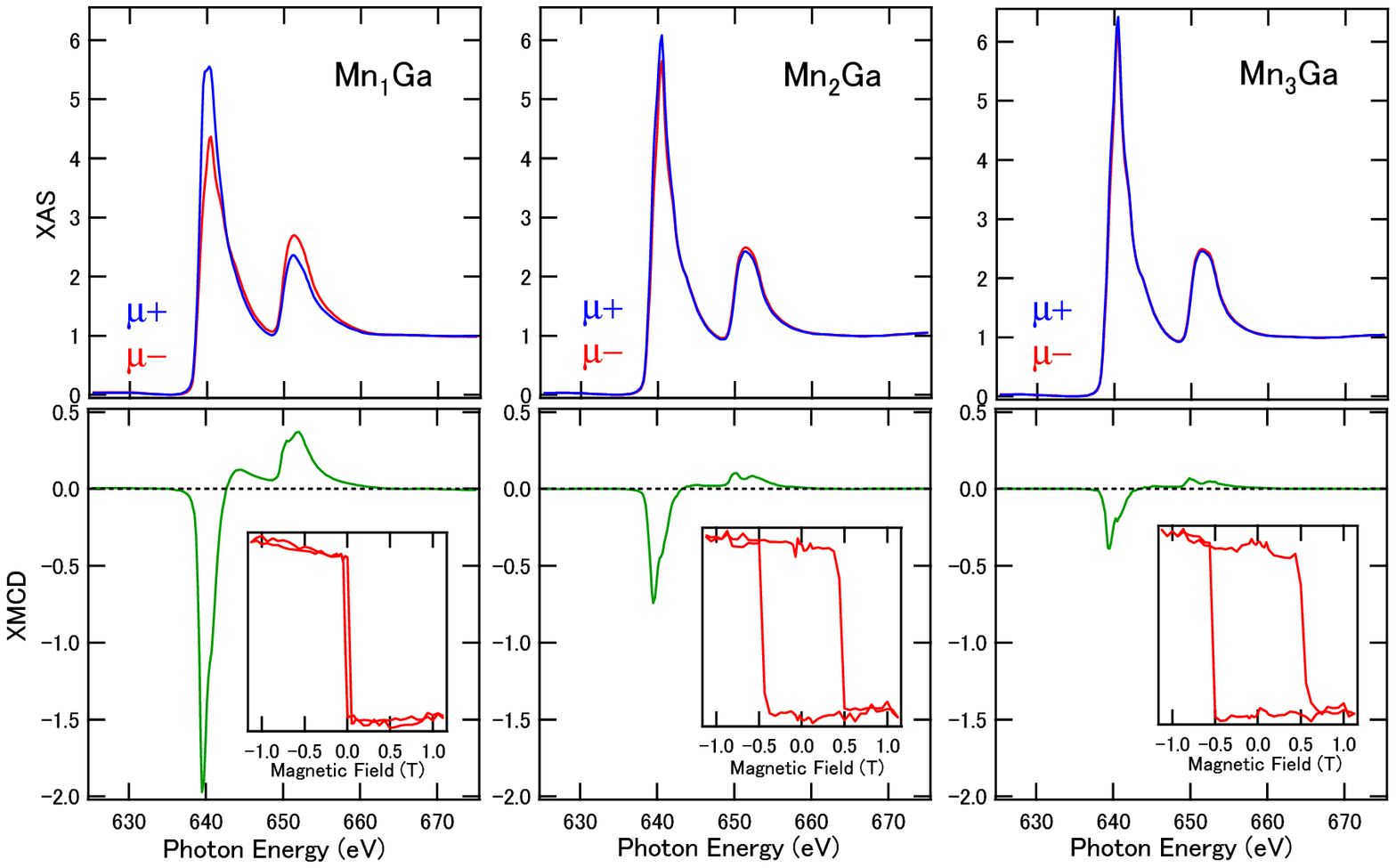}%
\end{center}
\caption{{\bf{XAS and XMCD of Mn$_{3-{\delta}}$Ga for ${\delta}$ = 0, 1, and 2.}} Spectra were measured at the normal incident setup where the incident beam and magnetic field were parallel to the sample film normal. ${\mu}^+$ and ${\mu}^-$ denote the absorption in different magnetic field direction. The insets show the magnetic field dependence of the hysteresis curves taken by fixed $L_3$-edge photon energy. All measurements were performed at RT. }
\end{figure}

\begin{figure}
[ptb]
\begin{center}
\includegraphics[
height=8in,
width=6in
]%
{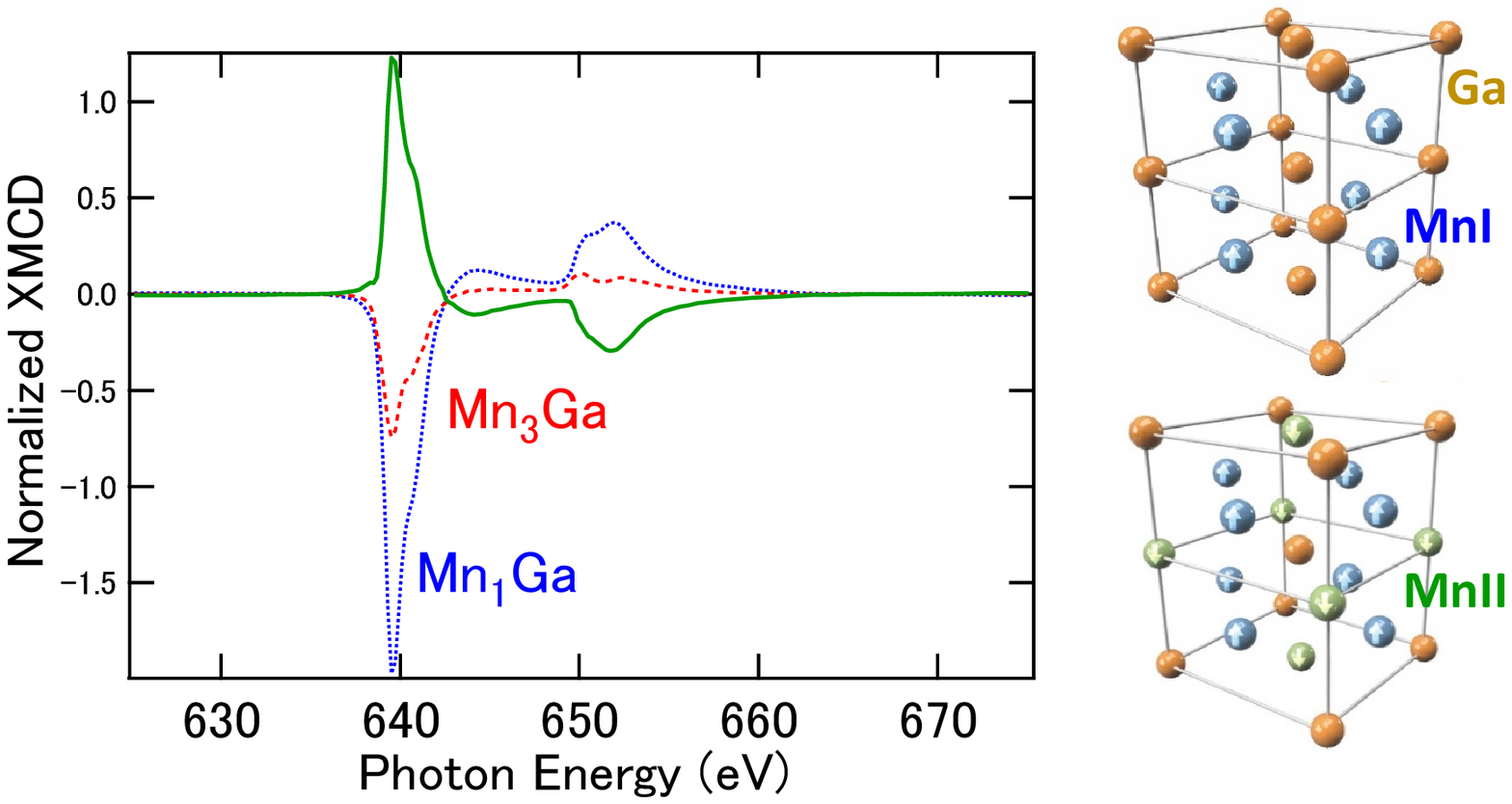}%
\end{center}
\caption{{\bf{Deconvoluted XMCD spectra of Mn$_{3-{\delta}}$Ga by subtraction from Mn$_1$Ga.}} The MnI and MnII components were separated in this procedure. Illustrations of the unit-cell structures of Mn$_1$Ga and Mn$_3$Ga are also displayed. }
\end{figure}

\begin{figure}
[ptb]
\begin{center}
\includegraphics[
height=8in,
width=6in
]%
{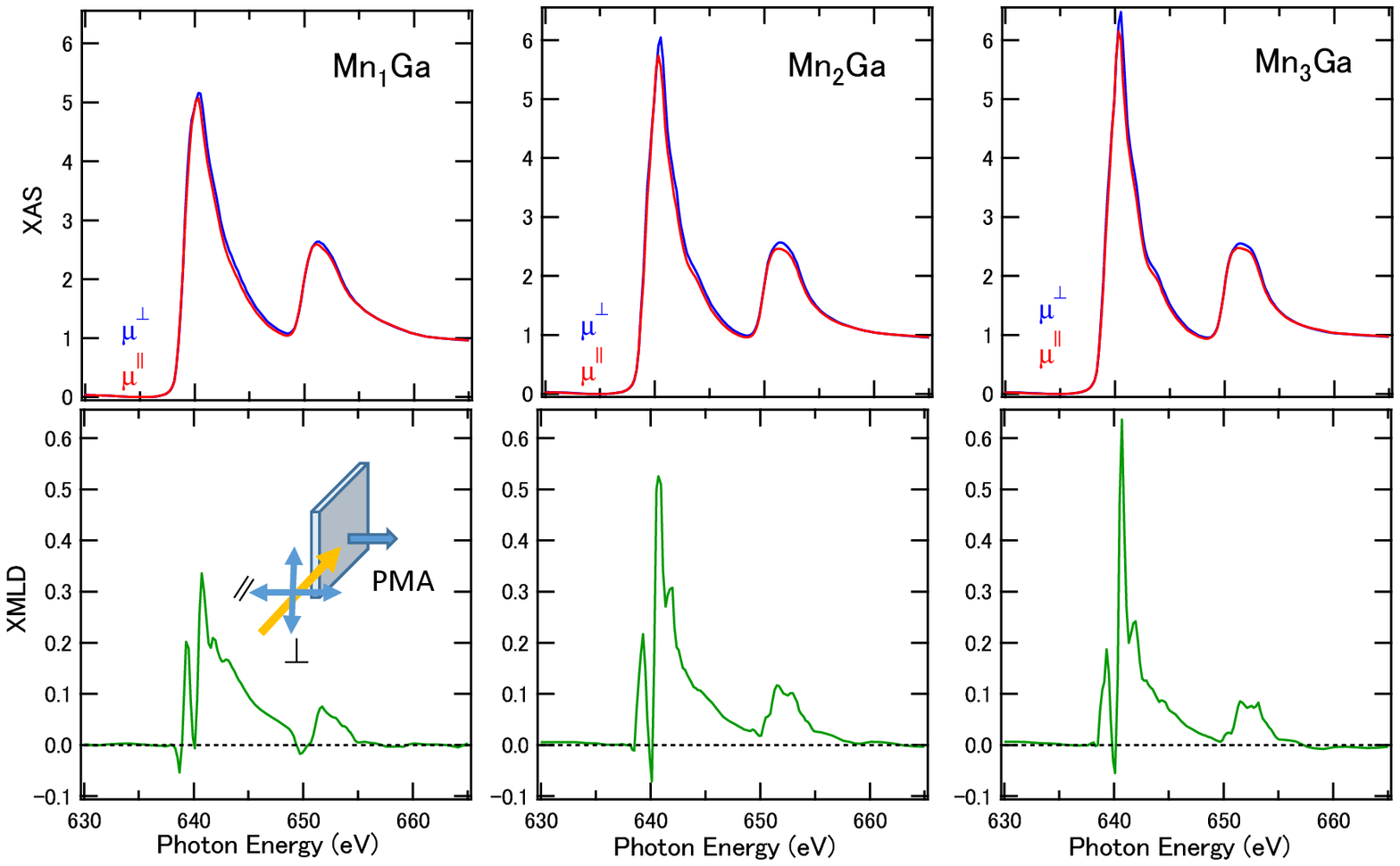}%
\end{center}
\caption{{\bf{XAS and XMLD of Mn$_{3-{\delta}}$Ga for  ${\delta}$ = 0, 1, and 2.}} Spectra were taken at the grazing incident setup where $\bf{E}$ of the incident beam and direction of magnetization $\bf{M}$ were parallel and perpendicular, respectively.  ${\mu}^{\perp}$ and ${\mu}^{\parallel}$ denote the absorption in different electric-field directions. The inset shows an illustration of the XMLD measurement geometry. The angle between sample surface normal and incident beam is set to 60$^{\circ}$. All measurements were performed at RT. }
\end{figure}

\begin{figure}
[ptb]
\begin{center}
\includegraphics[
height=8in,
width=6in
]%
{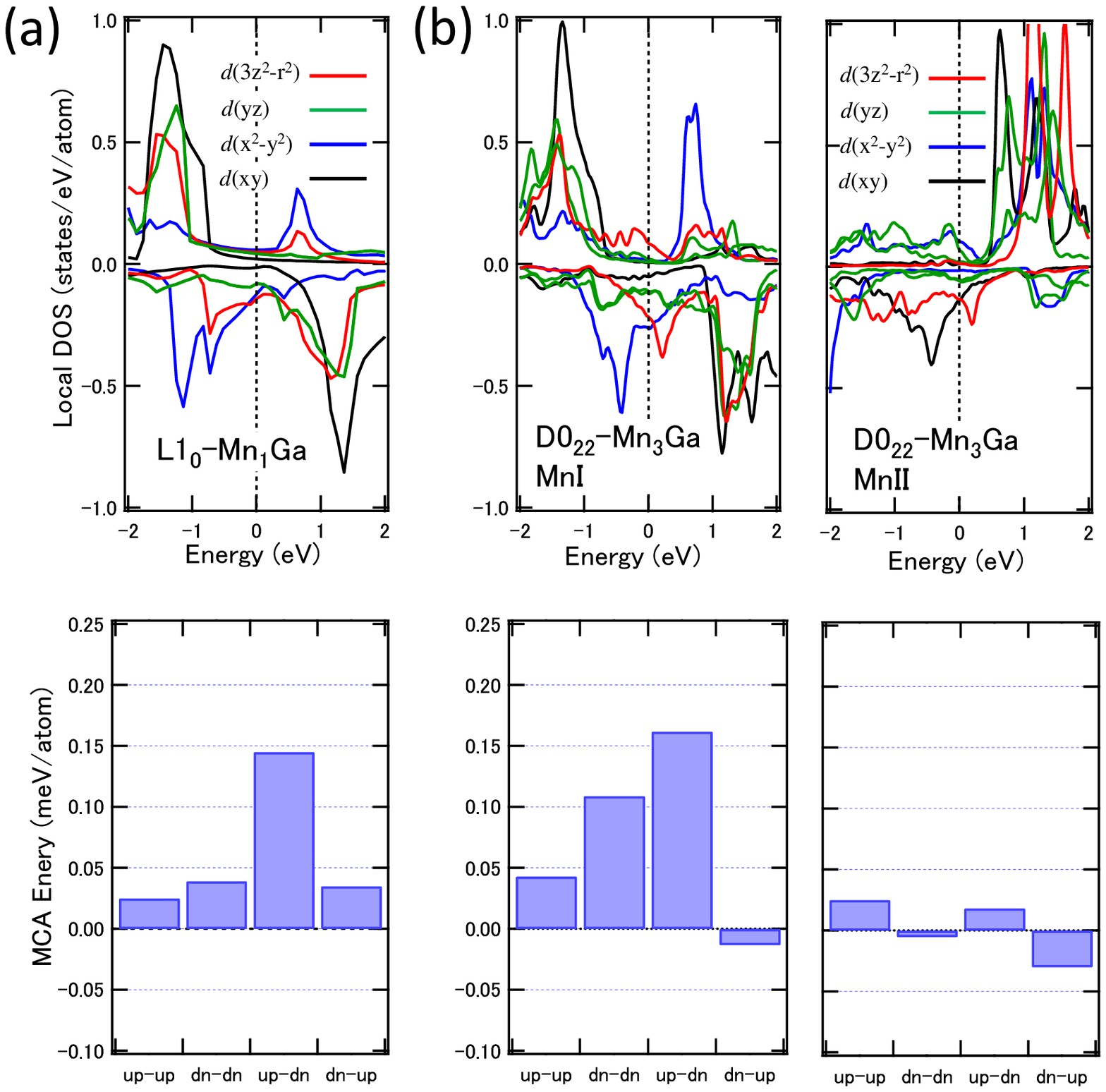}%
\end{center}
\caption{{\bf{DFT calculation of Mn$_{3-{\delta}}$Ga.}} Density of states of MnI and MnII components for each 3$d$ orbital state (upper panel) and bar graphs of the second-order perturbative channels of the spin-orbit interaction to the magneto-crystalline anisotropy energy (lower panel): 
(a) Mn$_1$Ga for MnI component and (b) Mn$_3$Ga for MnI and MnII components. }
\end{figure}

\end{document}